\documentclass{aa}

\usepackage{graphicx}
\usepackage{txfonts}
\usepackage{natbib}
\bibpunct{(}{)}{;}{a}{}{,}

\begin{document}

\title{The donor star of the X-ray pulsar X1908+075\thanks{Based on observations 
made with the William Herschel Telescope operated on the island of La Palma by the 
Isaac Newton Group in the Spanish Observatorio del Roque de los Muchachos of the 
Instituto de Astrof\'{\i}sica de Canarias.}}

\author{S. Mart\'{\i}nez-N\'u\~{n}ez\inst{1}
\and
A. Sander\inst{2}
\and
A. G\'{\i}menez-Garc\'{\i}a\inst{1,3}
\and
A. G\'onzalez-Gal\'an\inst{2}
\and
J.~M.~Torrej\'on\inst{1,3}
\and
C. G\'onzalez-Fern\'andez\inst{4}
\and
W.-R. Hamann\inst{2}
}
\offprints{S. Mart\'{\i}nez-N\'u\~{n}ez}
\institute{X-ray Astronomy Group. Departamento de F\'{\i}sica, Ingenier\'{\i}a de Sistemas y Teor\'{\i}a de la Se\~{n}al, 
University of Alicante, P.O. Box 99, E03080 Alicante, Spain. \email{silvia.martinez.nunez@gmail.com}
\and
Institut f\"ur Physik und Astronomie, Universit\"at Potsdam,
Karl-Liebknecht-Str. 24/25, D-14476 Potsdam, Germany
\and Instituto Universitario de F\'isica Aplicada a las Ciencias y las Tecnolog\'ias, University of Alicante, 
P.O. Box 99, E03080 Alicante, Spain
\and
Institute of Astronomy, University of Cambridge, Madingly
Road, Cambridge, CB3 0HA, UK}

\date{Received <date>; accepted <date>}
\abstract{

  High-mass X-ray binaries (HMXBs) consist of a massive donor star and a compact object.
  While several of those systems have been well studied in X-rays, little is known for
  most of the donor stars as they are often heavily obscured in the optical and ultraviolet
  regime. There is an opportunity to observe them at infrared wavelengths, however. The goal of 
  this study is to obtain the stellar and wind parameters of the donor star
  in the X1908+075 high-mass X-ray binary system with a stellar atmosphere model to
  check whether previous studies from X-ray observations and spectral morphology lead to
  a sufficient description of the donor star.
  We  obtained H- and K-Band spectra of X1908+075 and analysed them with the 
  Potsdam Wolf-Rayet (PoWR) model atmosphere code. For the first time, we  
  calculated a stellar atmosphere model for the donor star, whose main 
  parameters are: $M_{\mathrm{spec}}$ = $15 \pm 6 $\,$M_{\odot}$, $T_{\ast} = 23_{-3}^{+6},$kK, 
  $\log g_\mathrm{eff} = 3.0 \pm 0.2$ and $\log L/L_{\odot}$ = $4.81 \pm 0.25$. The obtained 
  parameters point towards an early B-type (B0--B3) star, probably in a supergiant phase. Moreover 
  we  determined a more accurate distance to the system of 4.85 $\pm$ 0.50 kpc than 
  the previously reported value.
}


\maketitle
%

\section{Introduction}
  \label{sec:intro}

High-mass X-ray binaries (HMXBs) are composed of a compact star orbiting a donor star from which 
there is an accretion of material \citep[see][for a review]{chaty2013}. There is a broad literature 
of the X-ray studies of these types of systems compared to the few detailed studies performed about 
the donor stars using stellar atmosphere models \citep[e.g.][]{Clark02,AGG14}.  The physical 
characteristics of the donor stars harbouring high-mass X-ray binaries, however, are essential for
a complete picture of the physical processes occurring in these binary systems.

The binary system X1908+075, also known as 4U 1909+07, was first discovered in 1978 by 
\citet{Forman78} with the Uhuru satellite. The source has been observed in surveys carried out 
with OSO7, Ariel, HEAO-1, EXOSAT, and INTEGRAL satellites, among others. It is a persistent 
X-ray source that shows fluctuations of 10$\%$ in the soft X-rays, 2--12 keV 
energy range \citep{Levine04}.

The binary system is a highly absorbed and faint pulsar that shows strong X-ray pulsations 
at a period of 605\,s \citep{Levine04}. The location of X1908+075 in the pulse--orbital period 
plane \citep{Corbet86} clearly indicates that this is a wind-fed, high-mass X-ray binary.

\citet{Levine04} determined the orbital parameters of the system (see Table~\ref{tab:orb}) 
with {\it RXTE/PCA} pointed observations. This study concludes that the system contains a highly 
magnetized neutron star orbiting in the wind of a massive companion star (mass function of 
(6.07 $\pm$ 0.35)M$_\odot$) with an orbital period of $4.400\pm 0.001$ days. Thus, they identify 
X1908+075 as a high-mass X-ray binary. Moreover, this study found changes in the optical depth 
along the line of sight to the compact object correlated with the orbital phase. These changes 
could be causing the observed soft X-ray modulation.

\citet{Levine04} estimate a mass of the donor star in the range of 9--31\,M$_\odot$ and 
an upper limit on its size of about 22\,R$_\odot$, using the mass function and the orbital 
inclination angle. Previous values were derived from  modelling the orbital phase-dependence of 
the X-ray flux. They also infer a wind mass-loss rate for the donor star of 
$\gtrsim$ $4 \times 10^{-6}$ M$_\odot$ yr$^{-1}$, which is larger than the expected theoretical 
value according to \citet{Vink2000}. Given the high rate combined with the estimated mass and 
radius, \citet{Levine04} concluded that the system might be a Wolf-Rayet star with a neutron 
star companion that could evolve and become a black hole-neutron star system 
in 10$^{4}$ to 10$^{5}$\,yrs.

\citet{Morel05} analysed near-infrared observations of stars in, or close to, the error box of 
\textit{HEAO-1/A3}. They suggest that the optical counterpart of the system might be a late 
O-type supergiant at a distance of 7 $\pm$ 3 kpc.

In this paper, we present intermediate-resolution ($R\sim 2500 $) infrared spectroscopic 
observations of the counterpart of the X-ray pulsar X1908+075 in the error box reported 
by the Chandra position. For the first time, a non-LTE stellar model atmosphere
analysis of X1908+075 is performed providing important physical parameters, 
such as the stellar temperature ($T_{\ast}$), the terminal velocity of the wind ($\varv_{\infty}$)
and the effective surface gravity ($g_\text{eff}$), for the donor star. We also calculate a 
more accurate value of the distance to the binary system and enclose the value of its inclination. 
We conclude that the optical counterpart is, in contrast to previous assumptions, an early 
B-type (B0--B3) star, probably in a supergitant phase.

In Sect.\,\ref{sec:obs} we describe the details of the observations, followed by a short
overview of the stellar atmosphere models in Sect.\,\ref{sec:models}. The results are shown in 
Sect.\,\ref{sec:results}, discussed in Sect.\,\ref{sec:discus}, and conclusions
are given in Sect.\,\ref{sec:conclu}.


\begin{table}
  \caption{Relevant orbital parameters derived by \citet{Levine04}}
  \label{tab:orb}

  \centering
  \begin{tabular}{c c}
                \hline\hline
                      Parameter       & Orbital second-epoch Values \rule[0mm]{0mm}{3mm}\\
                \hline
    $a_x$ $\sin i$ \rm{[cm]}    & (1.43 $\pm$ 0.03) $\times$ 10$^{12}$ \rule[0mm]{0mm}{3mm}\\
    $\tau_{90}$ \rm{[MJD]}     & 52631.383  $\pm$ 0.013\\
    $P_{\rm spin}$ ([\rm s])                 & 604.684  $\pm$ 0.001\\
    $\dot{P_{\rm spin}}$ [\rm{s} \rm{s}$^{-1}$]  & (1.22 $\pm$ 0.09) $\times$ 10$^{-8}$\\
    $e$                                   & 0.021 $\pm$ 0.039\\ 
    $P_{\rm orb}$ \rm{[days]}      & 4.4007 $\pm$ 0.0009\\
    $f(M) [M_{\odot}]$      & 6.07 $\pm$ 0.35\\
                \hline
  \end{tabular}

\end{table}



\section{Observations}
  \label{sec:obs}

In July 2009 (proposal id: 135-WHT45/09A), we obtained two intermediate-resolution 
spectra (K and H-Band spectra at 55018.97 MJD and 55019.92 MJD respectively) of the 
optical counterpart of X1908+075 using the Long-slit Intermediate Resolution 
Infrared Spectrograph (LIRIS) mounted on the 4.2\,m William Herschel Telescope (WHT), at 
the Observatorio del Roque de los Muchachos (La Palma, Spain). The instrument is equipped with a 
1024 $\times$ 1024 pixel HAWAII detector. We took advantage of the excellent seeing and made use of 
the $0.65\arcsec$ slit in combination with the intermediate-resolution K and H pseudogrisms. 
The K configuration covers the 2055$-$2415 nm range, giving a minimum resolving power 
$R$ $\sim$ 2500 at 2055 nm and a slightly higher resolving power at longer wavelengths. 
The H configuration covers the 1520$-$1783 nm range, giving a minimum R $\sim$ 2500 at 1520 nm.

\begin{figure}[th]
  \centering
  \includegraphics[width=0.4\textwidth]{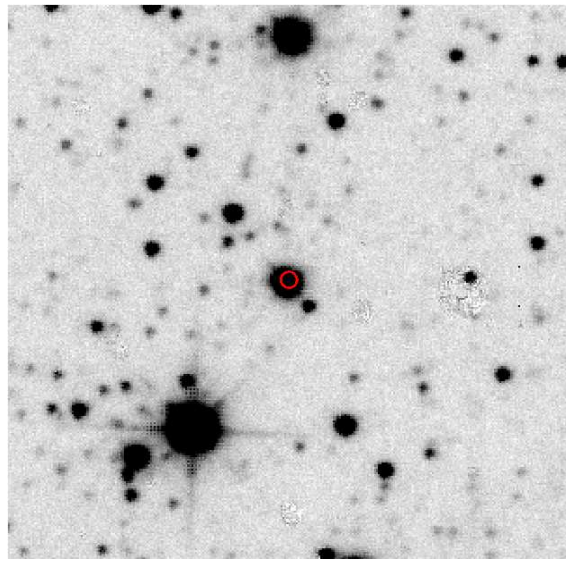}
  \caption{UKIDSS image of the X1908+075 region. North is up and east is left. The red circle indicates
the Chandra position of the X-ray source. Size of the image: 1\arcmin $\times$ 1\arcmin.}
  \label{fig:finder}
\end{figure}

The position of the X-ray source was accurately determined  using the High Energy Transmission Gratings 
(HETG) on board  {\it Chandra}, as the intersection between grating orders $m=\pm 1$ and the zero 
order image. The coordinates of the X-ray source are $\alpha=19^\mathrm{h} 10\arcmin 48.2\arcsec$ and 
$\delta=+07^{\deg} 35\arcmin 51.8\arcsec$, with an estimated uncertainty of $0.65\arcsec$. We then 
searched for appropriate candidates within the error circle using the UKIDSS database. In 
Fig.~\ref{fig:finder}, we present the K-Band image with the {\it Chandra} error circle overlay. 
Only one suitable candidate is compatible with the error circle, identified as 
2MASS J19104821+0735516, in agreement with candidate A in \citet{Morel05}.

\begin{figure*}[th]
  \centering
  \includegraphics[width=17cm]{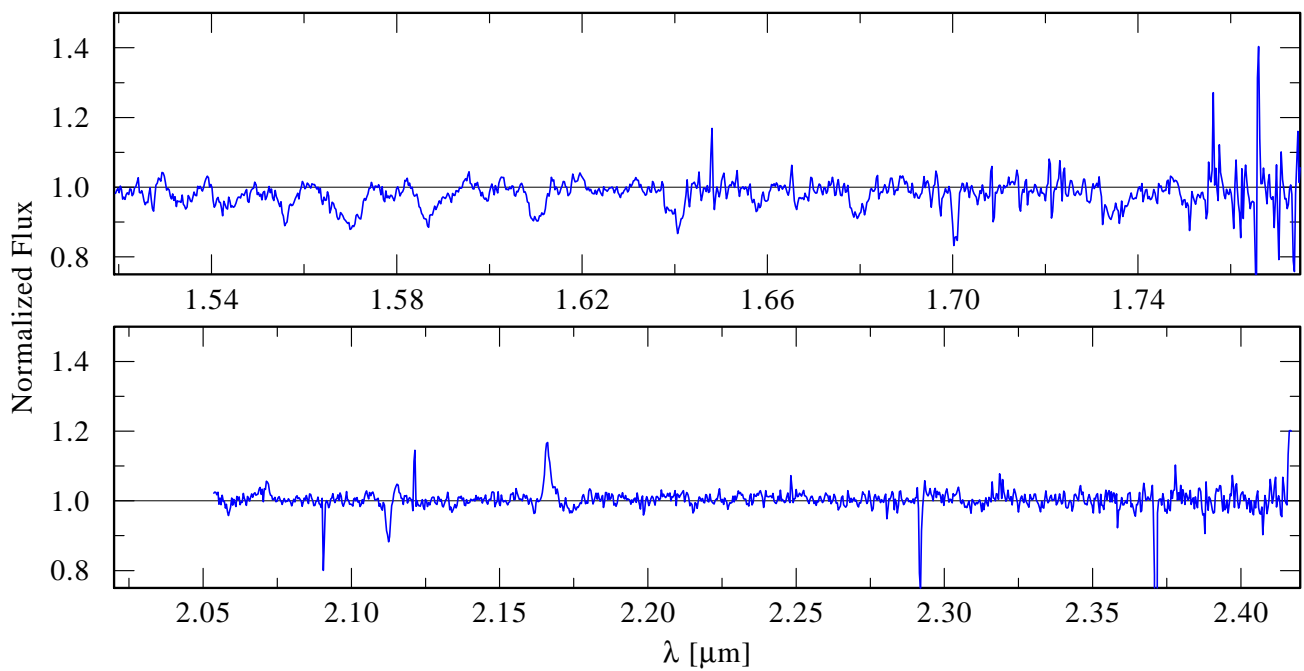}
  \caption{H and K-Band spectra of the X1908+075 donor star.}
  \label{fig:rawspec}
\end{figure*}

Data reduction was carried out using dedicated software developed by the LIRIS science group, 
which is implemented within IRAF\footnote{IRAF is distributed by the National Optical Astronomy 
Observatories, which are operated by the Association of Universities for Research in Astronomy, 
Inc., under cooperative agreement with the National Science Foundation}. We use the G2V star 
HD182081 and the A0V star BD-013649B to remove atmospheric features using the IRAF standard 
procedure. The normalized spectra are shown in Fig.\,\ref{fig:rawspec}.

\section{Stellar atmosphere models}
  \label{sec:models}

During the analysis, the observed spectra are compared with synthetic spectra calculated
with the Potsdam model atmosphere code (PoWR). The PoWR code provides a 
model for a spherical symmetric star with an expanding atmosphere by 
iteratively solving the radiative transfer equation and the statistical equilibrium equations 
in non-LTE. The code further includes energy conservation and
treats both the wind and  photosphere in a consistent scheme.

The main aspects of the code are summarized in \citet{GKH2002} and \citet{HG2003}.
This code has been applied to all kinds of stars that could be potential donors in 
wind-fed HMXB systems, such as O and B stars \citep[e.g.][]{Oskinova+2011,Evans+2012} 
and Wolf-Rayet (WR) stars \citep[e.g.][]{HGL2006,SHT2012}. Wind inhomogeneities are 
considered in a so-called ``micro-clumping'' approach, assuming that the wind is not smooth, 
but instead consisting of optical thin cells with an increased density and a void 
interclump medium \citep[cf.][]{HK1998}.

The ``stellar radius'' $R_{\ast}$ marks the lower boundary of the model atmosphere 
and is set at a Rosseland optical depth of $\tau = 20$, where we assume that the hydrostatic 
equation is fulfilled. In the quasi-hydrostatic regime we use this equation to
calculate a hydrodynamically-consistent velocity field\citep{Sander+2015}. 
In the outer part we assume a so-called ``beta-law','
\begin{equation}
  \label{eq:betalaw}
  \varv(r) = \varv_\infty \left( 1 - \frac{R_{\ast}}{r} \right)^{\beta}
,\end{equation}
with $\varv_\infty$ being the terminal velocity of the wind from the donor star.
Both parts are connected such that a smooth transition is guaranteed. That means 
not only the velocities but  their gradients are equal at the connection point,
which is usually close to the sonic point where the wind velocity is equal to the
speed of sound. In the final model for the donor star of X1908+075, we adopted an
exponent of $\beta = 1.2$, after testing models with different 
$\beta$ values. Models with lower values of $\beta$ down to $0.6$ lead
to slightly lower fit qualities. The precise choice of beta has hardly 
any effect on the obtained parameters, for instance, for the estimated mass-loss rate: 
reducing $\beta$ by $0.2$ requires approximately an increase of $0.05\,$dex in $\dot{M}$ to 
still reproduce the observed strength of the Br$\gamma$ line.

With a given luminosity $L$, the stellar temperature $T_{\ast}$ can be obtained from
$L$ and $R_{\ast}$ via the Stefan-Boltzmann relation. Apart from $L$, $T_{\ast}$, and 
$\varv_{\infty}$, the other main model parameters are the effective surface gravity 
$g_\text{eff}$ and the wind mass-loss rate $\dot{M}$. The $\log g_{\mathrm{eff}}$ corrects for full 
radiation pressure $a_\text{rad}$, i.e. it considers not only Thomson scattering, but
also the non-negligible continuum and line accelerations. While the actual
calculation for the model stratification is depth dependent, we use a mean value of 
$\Gamma_\text{rad} := a_\text{rad} / g$ to define an output value for
$g_\text{eff}$ that is related to the pure gravitational acceleration via
\begin{equation}
  \log g_\text{eff} = \log g + \log \left( 1 - \overline{\Gamma}_\text{rad} \right)\text{.}
\end{equation}
The details are described in \citet{Sander+2015}.
In addition, the clumping factor $D$ describes the maximum density contrast reached in the outer parts of the wind. In our models we
assume that the density contrast starts around the sonic point and increases outwards
until the specified value of $D$ is reached. For our O- and B-star models, we assume a solar composition according to \citet{Asplund2009}, see 
Table~\ref{tab:fitparam}. An overview of the model atoms used in the 
PoWR calculations is given in Table~\ref{tab:modelatom}.


\begin{figure*}
  \centering
  \includegraphics[width=17cm]{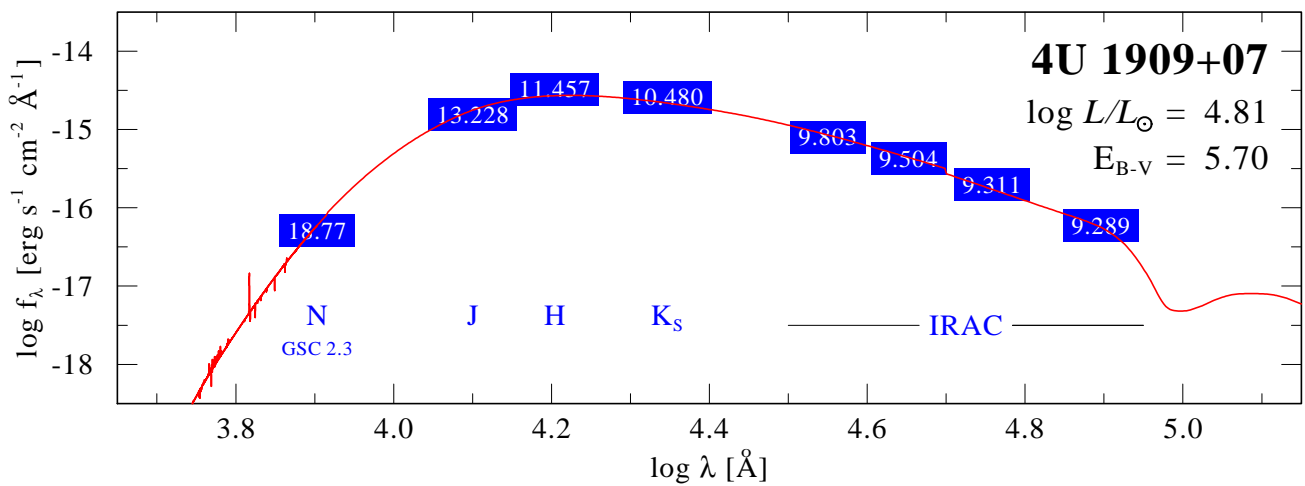}
  \caption{Spectral energy distribution (SED) of the 4U 1907+09 donor star. The PoWR model
           (red solid line) is reddened such that it fits the observed photometry marks (blue boxes).}
  \label{fig:sed}
\end{figure*} 

\section{Results}
  \label{sec:results}

\begin{figure}
 \centering
\includegraphics[angle=-90,width=0.5\textwidth]{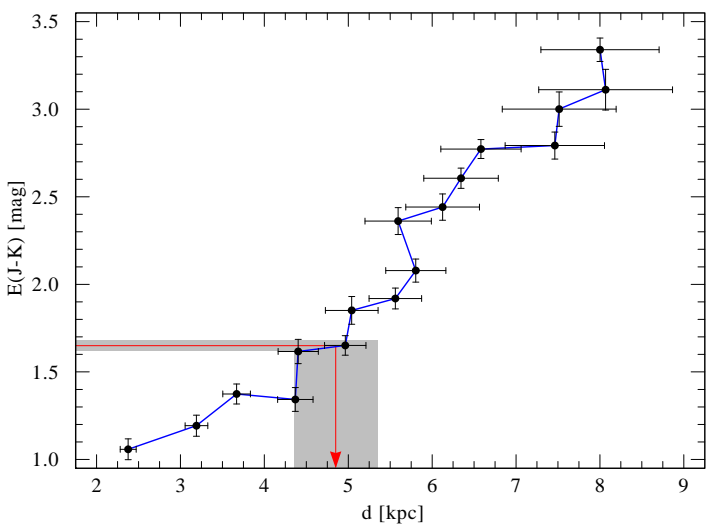}
\caption{Estimated extinction curve from the direction of X1908+075, obtained following the procedure outlined in \citet{CL2007}.}\label{fig:ext}
\end{figure}

We simultaneously reproduce the normalized line spectrum and the spectral energy distribution (SED) of 
X1908+075 by our calculated PoWR model (see Fig.~\ref{fig:sed}). The SED is compared to 
available photometry from IRAC \citep{IPAC2008}, 2MASS \citep{2MASS} and \citet{GSC2008}. First, 
the interstellar reddening colour excess $E_\text{B-V}$ is determined. Then, we apply the reddening
law by \citet{F1999} to calculate $E_\text{J-K}$ and compare this value with
the estimated extinction curve in the direction towards X1908+075. This curve, shown 
in Fig.~\ref{fig:ext}, was obtained following the procedure outlined in \citet{CL2007}: 
red clump giants are selected over a near-infrared colour-magnitude diagram. 
Since this population has a well-calibrated luminosity function, we can compare 
their intrinsic and apparent colour and magnitude to derive 
both their distance along the line of sight and the reddening to which they are subjected. 
Once this ($d$, $E_\text{J-K}$)-curve is established, we can compare it with the colour excess of our 
source ($E_\text{J-K} = 1.66$) to estimate a more accurate distance of 
4.85 $\pm$ 0.50 kpc compared to the previous estimation by \citet{Morel05}.

Since the total luminosity $L$ is critically dependent on the distance, we can estimate the luminosity, 
using our SED fit, once we have determined the new distance of the system. We obtain a value 
of $10^{4.81} L_{\odot}$. This value is used in our atmosphere model. As we already need 
a model for the SED fit, the whole process was iterated until we obtained the consistent solution described above.
The corresponding stellar radius $R_{\ast} = 16\,R_{\sun}$ inferred from 
temperature and luminosity perfectly agrees with the radius deduced for
a wind-fed system in \citet{Levine04}.

Prior to modelling the spectrum, we  compared our data with the 
standard atlases of OB stars in the $H$ and $K$ bands \citep{Hanson1998,Hanson2005}
to constrain the parameter space of the models. 
The \ion{He}{ii} lines at 1.693 $\mu$m and 2.188 $\mu$m are weak or maybe even 
absent as it is hard to distinguish them from the noise here. Together 
with the relatively strong hydrogen absorption lines in the H-band, this points to 
a spectral type later than O9.5 \citep{Hanson96,Hanson2005}. The luminosity class 
is more difficult to ascertain. Several indicators point to a high luminosity: 
The He I series lines are narrow and the H series can be seen, at our resolution, up to Br 19. However, 
the equivalent width of the He I line at 1.700 $\mu$m ($\sim 1.9 \AA$) is lower but comparable 
to that of the Brackett (11-4) line at 1.6814 $\mu$m (2.8 \AA). This is typically seen 
in supergiants rather than in main-sequence stars where EW(HeI) $ < $ EW(Br 11-4). All these 
properties suggest a high luminosity class. In addition, Br$\gamma$ in emission is a spectroscopic
feature only observed in two objects within the sample analysed by \citet{Hanson2005}. Both sources
belong to the rare class of ON/BN supergiants.

\begin{figure*}
  \centering
  \includegraphics[width=0.9\textwidth]{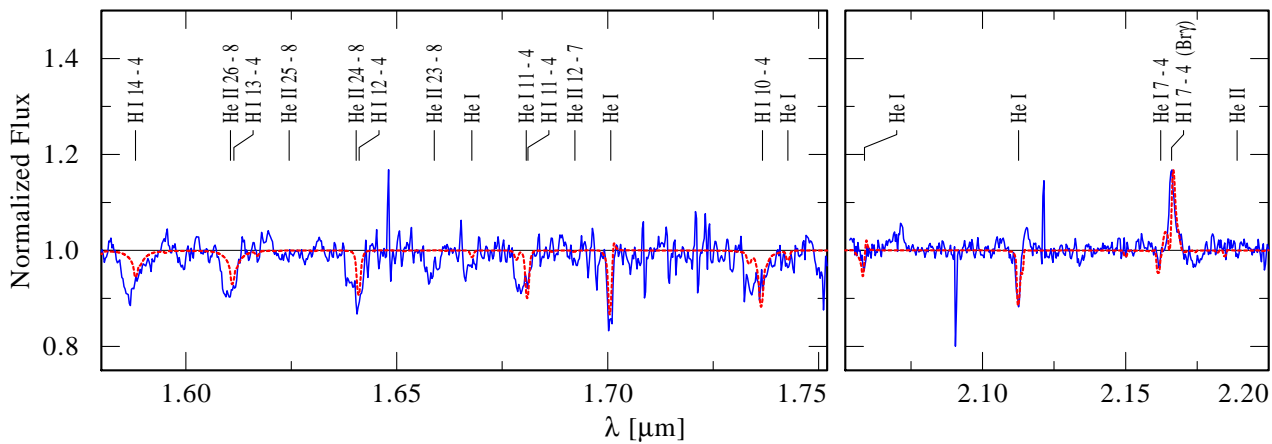}
  \caption{The observed spectrum (thin blue line) -- $H$ band (left panel), $K$ band (right panel) -- is compared with the best-fitting
     PoWR model (red thick dotted line). Noticeable spectral lines are identified.}
  \label{fig:normspec}
\end{figure*}

We calculate a set of stellar models starting with typical properties of O9.7-B4 stars.
An iterative improvement of the parameters is performed until a sufficient fit of both, the spectral 
lines and the photometry, is achieved. Figure~\ref{fig:normspec} shows the normalized line spectrum in the H and 
$\text{K}_\text{S}$ bands together with the best-fitting PoWR model. Most spectral features can 
be reproduced, apart from the observed asymmetric shape in some of the hydrogen absorption 
lines in the H band. The corresponding stellar parameters are compiled in Table\,\ref{tab:fitparam}. Because of the limited spectral range and the absence of unblended metal lines, there is 
a certain degeneracy in the spectral appearance that introduces a quite large error margin.
The observation is best reproduced by a model with $T_{\ast} = 23$\,kK and $\log g_\mathrm{eff} = 3.0$,
which is in line with an early B-type donor. To constrain the temperature, we used the Bracket series 
as well as \ion{He}{i} at 1.7 $\mu$m and \ion{He}{ii} at 1.693 $\mu$m. The absence
of \ion{He}{ii} emission at 2.189 $\mu$m excludes temperatures higher than $\approx 30\,$kK. In the
range between $21$ and $28\,$kK one can always find a model that reproduces Br$\gamma$ and a few other
lines sufficiently, but most of them fail to predict the strength of the very sensitive \ion{He}{i} 2.059
$\mu$m line. As the low terminal velocity deduced from Br$\gamma$ already favours the lower part
of the deduced temperature range, we found that a model with $T_{\ast} = 23$\,kK best fits 
the observation.

For $\log g_\mathrm{eff}$, the right wings of the bracket series could be used to determine an 
upper limit and excluded values $\ga 3.3$. These stronger effective gravity values 
would lead to broader wings than observed, including an unobserved red absorption wing for Br$\gamma$.

To reproduce the shape of the observed line profiles, the emergent spectrum 
of the PoWR model was convolved with a rotational velocity of $100\,$km/s. Given the limited 
spectral quality this should be seen as just a rough approximation with a relatively large
uncertainty of $\pm 50\,$km/s.

Before the donor was detected, \citet{Levine04} speculated that the donor might be a WR star. With the 
first infrared spectra of the donor obtained by \citet{Morel05} it became clear, that this 
is most likely not the case. Instead, \citet{Morel05} claim that the donor star would be an 
O7.5-9.5If supergiant, based on the emission feature next to the \ion{He}{i} absorption 
line at $2.11\,\mathrm{\mu m}$. This feature is actually a complex structure 
consisting of \ion{He}{i}, \ion{N}{iii}, \ion{C}{iii}, and \ion{O}{iii} emission lines. 
The C, N and O components are originated from high excitation states, and since not all 
of them are covered in our atomic database, this peak cannot be reproduced 
with the PoWR model. While one might argue that this line could be an indicator of enhanced 
nitrogen, this should also affect \ion{N}{iii} $2.2471$/$2.2513\,\mathrm{\mu m}$. Since 
these lines are not detectable in the spectrum, we keep a solar chemical composition.

The Br$\gamma$ line is the only line in the observed spectra formed in the stellar wind. Therefore, 
the determination of the terminal wind velocity can only rely on fitting the wind broadening 
of this line. This fitting results in $\varv_\infty = 500 \pm 100$\,km/s with a relatively 
large error margin. Significantly larger values would over predict the 
emission line width or even lead to additional absorption features that 
are not observed. 

The precise B-subtype is hard to determine from the given observations. The relationship 
between the He I line at 1.700 $\mu$m and the Brackett (11-4) line at 1.6814 $\mu$m (the former being 
deeper than the latter) clearly shows that the donor should be B0.5 or even B0 
\citep[Fig.\,2 in][]{Hanson1998}. The obtained temperature of $23\,$kK, however, favours a slightly 
later subtype around B2. We therefore specify the spectral type of the X1908+075 donor star to be 
in the range between B0 to B3.

\begin{table}
\caption{Derived properties of the X1908+075 donor star.}
  \label{tab:fitparam}
  \centering
  \begin{tabular}{ll}
    \hline\hline
      Parameter     &  Obtained value  \rule[0mm]{0mm}{3mm}  \\
    \hline
    $T_{\ast}$            &  $23.0^{+6}_{-3}\,$kK              \\
    $T_{\frac{2}{3}}$     &  $22.7^{+6}_{-3}\,$kK  \\
    $\log L/L_{\odot}\tablefootmark{a}$ &  $4.81 \pm 0.25$ \\
    $E_\text{B-V}\tablefootmark{b}$     &    $5.7 \pm 0.1$     \\
    $d\tablefootmark{c}$  &  $4.85 \pm 0.50$\,kpc    \\
    $\varv_{\infty}$      &  $500 \pm 100$\,km/s  \\
    $R_{\ast}$            &  $16\,R_{\odot}$          \\
    $\log g_\text{eff}$   &  $3.0 \pm 0.2 $  \\
    $\log g$              &  $3.2 \pm 0.2$  \\
    $M_{\text{spec}}$     &  $15 \pm 6 $\,$M_{\odot}$  \\
    $\dot{M}$             &  $10^{-6.55}$\,$M_{\odot}/$yr \\
    $D\tablefootmark{d}$  &  $10$ \\  
    $\varv_{\text{dop}}$  &  $20$\,km/s  \\
    $\varv_{\text{rot}}$  &  $100 \pm 50$\,km/s      \medskip \\
    $M_\text{V}$          &    $  -5.29$     \\
    $X_\text{H}\tablefootmark{e}$          &    $  0.735$     \\
    $X_\text{He}\tablefootmark{e}$         &    $  0.256$     \\
    $X_\text{C}\tablefootmark{e}$          &    $  2.2 \cdot 10^{-3}$     \\
    $X_\text{N}\tablefootmark{e}$          &    $  6.1 \cdot 10^{-4}$     \\
    $X_\text{O}\tablefootmark{e}$          &    $  5.3 \cdot 10^{-3}$     \\
    $X_\text{Fe}\tablefootmark{e}$         &    $  1.1 \cdot 10^{-3}$     \\
    $\log Q_{0}\tablefootmark{f}$  &  $47.1\,$s$^{-1}$  \\
    \hline
  \end{tabular}
  \tablefoot{
     \tablefoottext{a}{Luminosity assuming a distance of 4.85 $\pm$ 0.50\,kpc.}
     \tablefoottext{b}{Obtained via SED fitting, using the reddening law from \citet{F1999}.}
     \tablefoottext{c}{Distance inferred from consistent reddening fit and extinction curve shown in Fig.\,\ref{fig:ext}.}
     \tablefoottext{d}{Maximum value used in a depth-dependent approach, starting at the sonic point and increasing outwards.}
     \tablefoottext{e}{Solar abundances, adopted from \citet{Asplund2009}.}
     \tablefoottext{f}{Number of hydrogen ionizing photons}.}
\end{table} 



\begin{table}
  \caption{Overview of the used model atoms.}
  \label{tab:modelatom}
  \centering
  \begin{tabular}{lcc}  
    \hline\hline
       Ion  \rule[0mm]{0mm}{3mm}      &  Levels  &   Lines \\
    \hline 
     \ion{H}{i} \rule[0mm]{0mm}{3mm}  &   22   &   595  \\
     \ion{H}{ii}                      &    1   &   325  \\
     \ion{He}{i}                      &   35   &     0  \\
     \ion{He}{ii}                     &   26   &   231  \\
     \ion{He}{iii}                    &    1   &     0  \\
     \ion{C}{i}                       &    2   &     0  \\
     \ion{C}{ii}                      &   32   &   703  \\
     \ion{C}{iii}                     &   40   &   630  \\
     \ion{C}{iv}                      &   21   &   703  \\
     \ion{C}{v}                       &    1   &     1  \\
     \ion{N}{i}                       &    1   &     1  \\
     \ion{N}{ii}                      &   38   &   496  \\
     \ion{N}{iii}                     &   36   &   780  \\
     \ion{N}{iv}                      &   38   &   210  \\
     \ion{N}{v}                       &    2   &     0  \\
     \ion{O}{i}                       &    2   &     1  \\
     \ion{O}{ii}                      &   36   &   630  \\
     \ion{O}{iii}                     &   33   &   528  \\
     \ion{O}{iv}                      &   25   &   300  \\
     \ion{O}{v}                       &    2   &     1  \\
     \ion{Fe}{i}\tablefootmark{a}     &    1   &     0  \\
     \ion{Fe}{ii}\tablefootmark{a}    &    2   &     1  \\
     \ion{Fe}{iii}\tablefootmark{a}   &    8   &    25  \\
     \ion{Fe}{iv}\tablefootmark{a}    &   11   &    49  \\
     \ion{Fe}{v}\tablefootmark{a}     &   13   &    69  \\
     \ion{Fe}{vi}\tablefootmark{a}    &   17   &   121  \\
     \ion{Fe}{vii}\tablefootmark{a}   &   11   &    52  \\
     \ion{Fe}{viii}\tablefootmark{a}  &   1    &     0  \\
    \hline 
  \end{tabular}
  \tablefoot{
     \tablefoottext{a}{A super-level approach is used for this element. Fe actually denotes a 
generic atom that also includes the additional iron group elements Sc, Ti, V, Cr, Mn, Co, and Ni.
The details and relative abundances are given in \citet{GKH2002}.}}
\end{table} 

\newpage

\section{Discussion}
\label{sec:discus}

\subsection{On the accretion scenario}

We are going to demonstrate that the wind mass-loss rate derived in our model 
is compatible with the observed average X-ray luminosity, considering a Bondi-Hoyle-Lyttleton
(BHL) accretion scenario \citep[][and references therein]{Bondi44}. We estimated the fraction 
of the wind that is gravitationally captured by the compact 
object to estimate the accretion 
luminosity. First, we determine the orbital separation of the neutron star to the donor star ($a$). 
Given our spectroscopic mass, the orbital period derived by \citet{Levine04} 
(see Table~\ref{tab:orb}) and assuming a canonical neutron star of 
1.4M$_{\odot}$, we obtain $a = (1.98 \pm 0.10) \times 10^{12}$\,cm. Second, we determine 
the Bondi-Hoyle accretion radius, and finally, we estimate the captured fraction of the wind as follows:

\begin{equation}
f(a) = \frac{\dot{M}_{\mathrm{captured}}}{\dot{M}_{\mathrm{wind}}} = \frac{r_{\mathrm{accr}}^2}{4a^2}  
\\ \mathrm{;} \\ r_{\mathrm{accr}} = \frac{2GM_{\mathrm{NS}}}{\varv(a)_{\mathrm{wind}}^2+\varv(a)_{\mathrm{orb}}^2}
.\end{equation}

Therefore, the captured fraction of the wind depends on the wind velocity and the 
orbital velocity as a function of the orbital separation. From our model we obtain 
a $\varv_{\mathrm{wind}} = 190 \pm 40 \; \mathrm{km/s}$ for the given $a$,
and we estimate an orbital velocity of $\varv_{\mathrm{orb}} = 330 \pm 20 \; \mathrm{km/s}$
(considering our spectroscopic mass, a canonical neutron star of 1.4M$_{\odot}$ 
and the obtained $a$), which leads to a captured fraction of the wind of 
$f(a) = 0.0044 \pm 0.0016$.

Once we know the captured fraction of the wind, we can estimate the X-ray luminosity generated 
by the accretion of this fraction of the wind onto the neutron star as follows:
\begin{equation}
L_{\mathrm{accr}} = \frac{GM_{\mathrm{NS}}\dot{M}_{\mathrm{captured}}}{R_{\mathrm{NS}}}\xi = \frac{GM_{\mathrm{NS}}f(a)\dot{M}_{\mathrm{wind}}}{R_{\mathrm{NS}}}\xi
,\end{equation}
where $\xi$ is the efficiency factor of energy conversion and $R_{\mathrm{NS}} = 15 \mathrm{km}$, 
obtaining a $L_{\mathrm{accr}} = [(10 \pm 3) \times 10^{36}] \; \xi \; \mathrm{erg/s}$. 

However, we know that the source is persistent in X-rays with an X-ray 
luminosity of $\sim (1-4) \times 10^{36} \; \mathrm{erg/s}$ in the $2-30$~keV band according to 
\citet{Levine04} and using our distance value. Since we expect significant
X-ray emission below 2\,keV, we estimate the X-ray luminosity in 
the 0.1--50 keV energy range assuming the spectral model used 
by \citet{Levine04}, obtaining $L_{\mathrm{X}}= (1.4-5.6) \times 10^{36} \; \mathrm{erg/s}$. 
Taking an average value of $L_{\mathrm{X}} = 3.5 \times 10^{36} \mathrm{erg/s,}$
we obtain an energy conversion efficiency of $0.36 \pm 0.13$, which is
a reasonable value for a wind-fed system.

\begin{figure}
  \centering
  \includegraphics[width=0.6\columnwidth,angle=90]{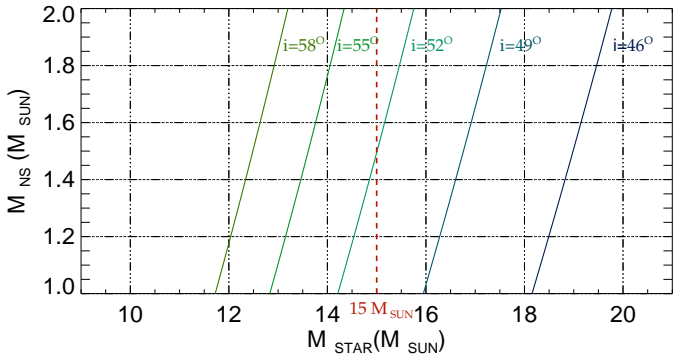}
  \caption{Relations between the donor mass and neutron star mass. Lines of constant orbital inclination constructed from the mass function of \citet{Levine04} are labelled. The 
  X-axis range corresponds to the spectroscopic mass error margin, while the Y-axis range covers the range permitted for a neutron star mass \citep[e.g.][]{Lattimer2012}.
}
  \label{fig:ana}
\end{figure} 

\subsection{On the wind mass-loss rate determination}

Our stellar atmosphere model is used to derive a wind mass-loss rate 
$\dot{M}_\text{PoWR} = 2.8 \times 10^{-7} \, M_{\odot}/$yr, significantly lower 
than the value of $\gtrsim$ $4 \times 10^{-6}$ M$_\odot$ yr$^{-1}$ derived
from X-rays by \citet{Levine04}. \citet{Levine04} employed the density stratification of the wind 
expected from a $\beta$-law of the velocity field and the continuity equation. Using the 
orbital parameters that they derived, they make a fit of the orbital modulation of the 
X-rays absorption ($N_\mathrm{H}$ ). From this fit it is possible to derive $\dot{M} / \varv_\infty$, 
so an assumption about $\varv_\infty$ is required to make an estimation of $\dot{M}$. Since our 
adopted $\varv_\infty$ is very low, it is very likely that \citet{Levine04} assumed a higher 
value of $\varv_\infty$. Therefore, their estimations of $\dot{M}$ must be corrected by a 
factor $F = 500 / \varv_\infty^\text{Levine}$. Even using this correction, it is clear that 
the estimations from the orbital modulation of the $N_\mathrm{H}$  yield higher values of $\dot{M}$ 
than the values obtained in this work. We have used a clumping factor $D=10$. Decreasing $D$ in 
order to increase $\dot{M}$, worsens the quality of the fit. For instance, a more modest 
value of clumping $D=4$ (see Appendix~\ref{sec:app}, in which models with different D values
are compared), yields a higher mass-loss rate $\dot{M}=4.6\cdot 10^{-7} \, M_{\odot}/$yr. In summary, 
there is a discrepancy between the X-rays estimation by \citet{Levine04} and our model 
estimation ($\dot{M}_\text{PoWR}$). The plausible reasons are: a density enhancement in the vicinity of the neutron star that systematically increases 
the measured $N_\mathrm{H}$ ; and/or an accumulation of uncertainties in both methods, including 
$\dot{M}$, $D$ for the stellar atmosphere calculations, and $\varv_\infty^\text{Levine}$, $N_\mathrm{H}$  
and its variability in  X-rays.

In addition to the comparison with the mass-loss rate estimation from X-rays, we 
can compare our estimation to the predictions by \citet{Vink2000} for OB stars. 
These predictions distinguish two regimes depending on the effective 
temperature of the star (the so-called \textit{\textup{bi-stability jump}}). These two regimes 
are also characterized by a different ratio $r = \varv_\infty / \varv_\text{esc}$, where $\varv_\text{esc}$ 
is the escape velocity (corrected for the radiative acceleration due to Thomson scattering). In the hot regime 
the effective temperature is over $30$\,kK and $r \sim 2.6$, whereas in the cool regime 
the effective temperature lies below $22.5$\,kK and $r \sim 1.3$. Hence, given our 
 parameters  (see Table~\ref{tab:fitparam}), and $\varv_\text{esc} = 560$~km/s, 
the donor star in X1908+075 belongs to the cool regime. \citet{Vink2000} gives an estimation 
of the mass-loss rate depending on the luminosity, the mass, the ratio $r,$ and the effective 
temperature of the star. Plugging our parameters  into the cool 
regime equation, we find $\dot{M}_\text{Vink} \simeq 8 \times 10^{-7} \, M_{\odot}/$yr.
The difference between our $\dot{M}_\text{PoWR}$ and the expected
$\dot{M}_\text{Vink}$ is on the order of $\sqrt{D}$, hinting that a 
porous wind treatment in the mass-loss predictions might 
possibly solve this discrepancy.

\subsection{Enclosing the inclination angle of the system}

We can also constrain the inclination of the system using our spectroscopic mass
($15 \pm 6 $\,$M_{\odot}$) and the mass function given by \citet{Levine04}. Figure~\ref{fig:ana} 
shows lines of constant orbital inclination for the allowed spectroscopic mass
range and the permitted neutron star mass range by the equation of nuclear
state \citep[e.g.][]{Lattimer2012}. The allowed inclination values are in the 
range 46$^{\circ}$  $<$ i $<$ 58$^{\circ}$. The canonical neutron star mass of 
1.4~M$_{\odot}$ favours an inclination of $\sim$ 50$^{\circ}$. Our determination 
agrees with the estimation of \citet{Levine04}, who used
a completely independent method and obtained the inclination from the orbital 
modulation of N$_H$.

\begin{figure}
  \centering
  \includegraphics[angle=270,width=0.85\columnwidth]{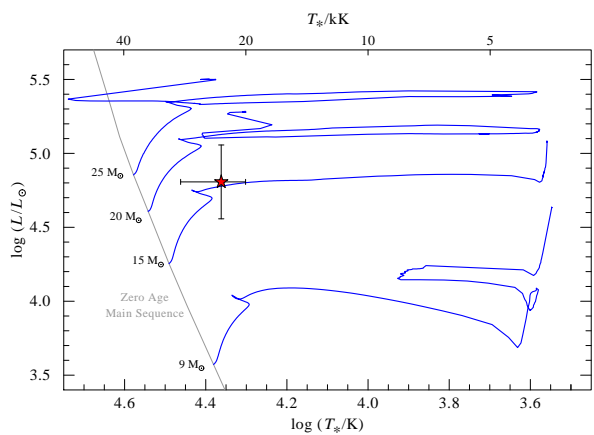}
  \caption{Evolutionary tracks using the Geneva Stellar Models \citep{Eks2012}
for solar metallicity and a rotation rate of $\varv_{\mathrm{ini}}/\varv_{\mathrm{crit}}=0.4$ on ZAMS
(it corresponds to 170 km/s for an initial mass of 20$M_{\odot}$). 
Initial masses are: 9, 15, 20, and 25$M_{\odot}$. The position of X1908+075
is marked by the red star.}
  \label{fig:HRD}
\end{figure} 

\subsection{On the evolutionary phase}

Our spectrum shows Br$\gamma$ in emission (see Sect.\,\ref{sec:results}), which 
indicates that the source could belong to the rare class of ON/BN supergiant systems. 
There are several observational hints that point to this classification:
\begin{itemize}

\item This class of objects commonly show variability \citep[e.g.][]{Walborn1976}. 
Comparing our data set with \citet{Morel05} reveals that Br$\gamma$ for X1908+075 is not 
only in emission, but  varies significantly between the two observations. 
We  measured an equivalent width for this line to be -4.3~\AA\ (55018.97 MJD), which 
differs by a factor of ~3 compared to the value of -1.4~\AA\ (52418.54 MJD) found by \citet{Morel05}. 
The peak height in the normalized spectrum is also about three times higher 
than in the observation discussed in \citet{Morel05}. 

\item  ON/BN supergiant systems seem to be nitrogen enriched \citep[e.g.][]{Corti2009}.
X1908+075 has likely gone through a common envelope phase in the past 
because of the proximity of the neutron star and its companion 
($R_\text{orb} \approx 1.7R_{\ast}$). Consequently, the source could be 
nitrogen enriched \citep[e.g.][]{Langer2012} and thus, 
show spectral features similar to ON/BN stars as seen 
in \citet{Hanson2005}. However, our spectrum does not show unblended 
metal lines, which makes it hard to determine the CNO abundances. Models with 
increased nitrogen did not improve the fit and thus we cannot confirm if 
the star is really nitrogen enriched or not.

\item ON/BN supergiant systems are candidates for interacting binaries 
\citep[e.g.][]{Walborn1976}. The projected rotational velocity 
derived from our data is high compared to that of isolated OB stars 
\citep{Simon-Diaz2014}. This could be the result of a
tight binary interaction of the system in the past. 

\item Figure~\ref{fig:HRD} shows the evolutionary tracks using the Geneva Stellar 
Models \citep{Eks2012} computed for single-star evolution and initial 
masses of $9$, $15$, $20$, and $25\,M_{\odot}$. Thus, assuming that binary 
interactions in our system have been negligible in the past, we can compare 
the current position of our donor star to these tracks. Indeed, the position 
of X1908+075 in the HR diagram is very close to the $15\,M_{\odot}$-track in 
coincidence with our spectroscopic mass. However, as previously discussed, 
there are several indications that point towards a former interaction that 
would make the star to appear much younger than an isolated star \citep{Langer2012}.
Nevertheless, we can neither confirm nor rule out this expected rejuvenation of 
the system due to the large uncertainties in our spectroscopic mass 
and effective temperature.
\end{itemize}

\subsection{On the asymmetry of some of the hydrogen absorption lines}

The hydrogen absorption lines of our data show an additional
component with respect to the model (see Fig,\,\ref{fig:normspec}). These 
additional absorptions have a measured blueshift of about 200 km/s compared 
to the rest of the spectrum. The additional absorption
is comparable to the intensity of the lines in the model. This effect is 
only seen in the hydrogen absorption lines, while the helium absorption 
lines and Br$\gamma$ are perfectly reproduced by our model atmosphere. 

A first hypothesis would be that these lines might be shifted because of the 
tidal disruption expected in this kind of close binary systems \citep{AMK04} 
in which the optical companion is pear-shaped, leading to a non-uniform 
temperature distribution in the atmosphere of the star. Even though 
we cannot completely rule out this hypothesis since PoWR assumes 
a spherically symmetric wind, it is certainly quite unlikely
since there is no reason why this effect is only seen in the hydrogen
absorption lines.

Another hypothesis is that these additional absorptions are due to the presence
of a density enhancement caused by the interaction with the neutron star, 
as seen in the simulations of \citet{Blondin91} and \citet{Mano2012}. 
This hypothesis is supported by the fact that these blueward shifts of the 
absorption features are only seen in the H-band spectrum, while 
the lines in the K-band spectrum are perfectly reproduced. Both spectra were 
taken at different orbital phases. The H-band spectrum was taken 
at orbital phase  $\phi=0.158$, with the neutron star travelling to quadrature where 
the projected orbital velocity towards the observer is maximum. For an orbital 
speed of $\varv_{\mathrm{orb}}=330~\mathrm{km/s}$ and an orbital inclination of the 
order of $i\approx 50^{\rm o}$ this component would be $\varv_{\mathrm{orb}}\sin i=252~\mathrm{km/s}$,
of the same order than the measured blueshift. Moreover, the K-band spectrum 
was taken during orbital phase $\phi=0.374$ with the neutron star approaching 
inferior conjunction where the projected orbital velocity towards the observer 
is zero. As a consequence, the K-band spectrum does not show this effect. 

 To reproduce the additional absorptions, however, we need the column density of
the extra absorber to be at least equal to the column density from the photosphere to the observer,
which is $\sim 1~\mathrm{g\,cm^{-2}}$ in our model, corresponding to 
$N_\text{H}~\sim ~60~\times~\mathrm{10^{22} at\,cm^{-2}}$. Moreover, this extra absorber has to be 
hot enough to produce a significant amount of hydrogen lines at level 4. \cite{Mano2012} 
simulations corresponding to our mass-loss rate and terminal velocity has shown a very 
modest density enhancement in the vicinity of the neutron star in comparison to our
required estimations. Furthermore, \citet{Levine04} were able to explain the observed 
$N_\text{H}$ orbital modulation appealing to the wind without any further structure. Therefore,
it is quite unlikely that the observed extra absorption is due to the presence of a density 
enhancement produced by the interaction of the neutron star with the stellar wind.

\section{Conclusions}\label{sec:conclu}

For the first time, we  obtained a reliable set of stellar and wind parameters for the donor star 
in the high-mass X-ray binary X1908+075, allowing us to determine various parameters of the system. 
Most of our results are in line with what is expected from a wind-fed binary system. 
The main conclusions of this work are:
\begin{itemize}
 
\item The donor star of X1908+075 is identified with an early B-type star (B0-B3). 
Its main parameters are: $M_{\mathrm{spec}}$ = $15 \pm 6 $\,$M_{\odot}$, $T_{\ast} = 23_{-3}^{+6}\,$kK, 
$\log g_\mathrm{eff} = 3.0 \pm 0.2$, and $\log L/L_{\odot}$ = $4.81 \pm 0.25$.

\item The low effective gravity perfectly fits with a supergiant star, favouring this classification. 
Moreover, the presence of Br$\gamma$ in emission could indicate that the source belongs
to the rare class of nitrogen-enriched B-type supergiant stars.

\item The source shows a variable stellar wind, since the equivalent width of Br$\gamma$ emission
line in our K-spectrum is three times higher than the value found by \citet{Morel05}.

\item The derived parameters along this work are consistent with a wind-fed high-mass X-ray binary within 
the Bondi-Hoyle-Lyttleton (BHL) accretion scenario.

\item The distance to the system is constrained to the value of 4.85 $\pm$ 0.50 kpc, which is well inside the 
previously estimated range of 7 $\pm$ 3 kpc reported by \citet{Morel05}.

\item The wind mass-loss rate we have obtained from the spectral fit is one order
of magnitude lower than the \citet{Levine04} estimation using X-ray observations.

\item The inclination of the system is constrained, favouring an inclination $\sim$ 50$^{\circ}$,
assuming a canonical 1.4~M$_{\odot}$ neutron star.
 
\end{itemize}

As a general conclusion, we would like to emphasize that  to understand
wind-fed binary systems in a coherent way, one needs to analyse the donor stars with 
stellar atmosphere models to constrain their parameters and to understand how 
the wind of the donor star interacts with its compact companion.

\begin{acknowledgements}

We thank the anonymous referee whose comments allowed us to improve this paper.
 
This work was supported by the Spanish Ministerio de Ciencia e Innovaci\'on through the 
project AYA2010-15431. SMN acknowledges the support of the Spanish Unemployment Agency, which allowed 
her to continue her scientific collaborations during the critical situation of the Spanish 
Research System. AS is supported by the Deutsche Forschungsgemeinschaft (DFG) 
under grant HA 1455/22. AGG acknowledges support by Spanish MICINN under 
FPI Fellowship BES-2011-050874 and the Vicerectorat 
d'Investigaci\'o, Desenvolupament i Innovaci\'o de la Universitat d'Alacant under
 project GRE12-35. AGG is supported by the Deutsches Zentrum f\"{u}r Luft und Raumfahrt (DLR) 
under contract No. FKZ 50 OR 1404. 

This publication makes use of data products from the Two Micron All Sky Survey and
UKIDSS project. 2MASS is a joint project of the University of Massachusetts and the Infrared 
Processing and Analysis Center/California Institute of Technology, funded by 
the National Aeronautics and Space Administration and the National Science Foundation.
The UKIDSS project is defined in \citet{Lawrence07}. UKIDSS uses the UKIRT Wide 
Field Camera (WFCAM; \citet{Casali07}). The photometric system is described in \citet{Hewett06}, 
and the calibration is described in \citet{Hodgkin09}. The pipeline processing 
and science archive are described in \citet{Lewis05} and \citet{Hambly08}. 

This publication was motivated by a team meeting sponsored by the International Space Science 
Institute at Bern, Switzerland.

\end{acknowledgements}

\bibliographystyle{aa}
\bibliography{4U1909}

\begin{thebibliography}{41}
\expandafter\ifx\csname natexlab\endcsname\relax\def\natexlab#1{#1}\fi

\bibitem[{{Abubekerov} {et~al.}(2004){Abubekerov}, {Antokhina}, \&
  {Cherepashchuk}}]{AMK04}
{Abubekerov}, M.~K., {Antokhina}, {\'E}.~A., \& {Cherepashchuk}, A.~M. 2004,
  Astronomy Reports, 48, 89

\bibitem[{{Asplund} {et~al.}(2009){Asplund}, {Grevesse}, {Sauval}, \&
  {Scott}}]{Asplund2009}
{Asplund}, M., {Grevesse}, N., {Sauval}, A.~J., \& {Scott}, P. 2009, \araa, 47,
  481

\bibitem[{{Blondin} {et~al.}(1991){Blondin}, {Stevens}, \&
  {Kallman}}]{Blondin91}
{Blondin}, J.~M., {Stevens}, I.~R., \& {Kallman}, T.~R. 1991, \apj, 371, 684

\bibitem[{{Bondi} \& {Hoyle}(1944)}]{Bondi44}
{Bondi}, H. \& {Hoyle}, F. 1944, \mnras, 104, 273

\bibitem[{{Cabrera-Lavers} {et~al.}(2007){Cabrera-Lavers}, {Hammersley},
  {Gonz{\'a}lez-Fern{\'a}ndez}, {L{\'o}pez-Corredoira}, {Garz{\'o}n}, \&
  {Mahoney}}]{CL2007}
{Cabrera-Lavers}, A., {Hammersley}, P.~L., {Gonz{\'a}lez-Fern{\'a}ndez}, C.,
  {et~al.} 2007, \aap, 465, 825

\bibitem[{{Casali} {et~al.}(2007){Casali}, {Adamson}, {Alves de Oliveira},
  {Almaini}, {Burch}, {Chuter}, {Elliot}, {Folger}, {Foucaud}, {Hambly},
  {Hastie}, {Henry}, {Hirst}, {Irwin}, {Ives}, {Lawrence}, {Laidlaw}, {Lee},
  {Lewis}, {Lunney}, {McLay}, {Montgomery}, {Pickup}, {Read}, {Rees}, {Robson},
  {Sekiguchi}, {Vick}, {Warren}, \& {Woodward}}]{Casali07}
{Casali}, M., {Adamson}, A., {Alves de Oliveira}, C., {et~al.} 2007, \aap, 467,
  777

\bibitem[{{Chaty}(2013)}]{chaty2013}
{Chaty}, S. 2013, Advances in Space Research, 52, 2132

\bibitem[{{Clark} {et~al.}(2002){Clark}, {Goodwin}, {Crowther}, {Kaper},
  {Fairbairn}, {Langer}, \& {Brocksopp}}]{Clark02}
{Clark}, J.~S., {Goodwin}, S.~P., {Crowther}, P.~A., {et~al.} 2002, \aap, 392,
  909

\bibitem[{{Corbet}(1986)}]{Corbet86}
{Corbet}, R.~H.~D. 1986, \mnras, 220, 1047

\bibitem[{{Corti} {et~al.}(2009){Corti}, {Walborn}, \& {Evans}}]{Corti2009}
{Corti}, M.~A., {Walborn}, N.~R., \& {Evans}, C.~J. 2009, \pasp, 121, 9

\bibitem[{{Ekstr{\"o}m} {et~al.}(2012){Ekstr{\"o}m}, {Georgy}, {Eggenberger},
  {Meynet}, {Mowlavi}, {Wyttenbach}, {Granada}, {Decressin}, {Hirschi},
  {Frischknecht}, {Charbonnel}, \& {Maeder}}]{Eks2012}
{Ekstr{\"o}m}, S., {Georgy}, C., {Eggenberger}, P., {et~al.} 2012, \aap, 537,
  A146

\bibitem[{{Evans} {et~al.}(2012){Evans}, {Hainich}, {Oskinova}, {Gallagher},
  {Chu}, {Gruendl}, {Hamann}, {H{\'e}nault-Brunet}, \& {Todt}}]{Evans+2012}
{Evans}, C.~J., {Hainich}, R., {Oskinova}, L.~M., {et~al.} 2012, \apj, 753, 173

\bibitem[{{Fitzpatrick}(1999)}]{F1999}
{Fitzpatrick}, E.~L. 1999, \pasp, 111, 63

\bibitem[{{Forman} {et~al.}(1978){Forman}, {Jones}, {Cominsky}, {Julien},
  {Murray}, {Peters}, {Tananbaum}, \& {Giacconi}}]{Forman78}
{Forman}, W., {Jones}, C., {Cominsky}, L., {et~al.} 1978, \apjs, 38, 357

\bibitem[{{Gonz{\'a}lez-Gal{\'a}n} {et~al.}(2014){Gonz{\'a}lez-Gal{\'a}n},
  {Negueruela}, {Castro}, {Sim{\'o}n-D{\'{\i}}az}, {Lorenzo}, \&
  {Vilardell}}]{AGG14}
{Gonz{\'a}lez-Gal{\'a}n}, A., {Negueruela}, I., {Castro}, N., {et~al.} 2014,
  \aap, 566, A131

\bibitem[{{Gr{\"a}fener} {et~al.}(2002){Gr{\"a}fener}, {Koesterke}, \&
  {Hamann}}]{GKH2002}
{Gr{\"a}fener}, G., {Koesterke}, L., \& {Hamann}, W. 2002, \aap, 387, 244

\bibitem[{{Hamann} {et~al.}(2006){Hamann}, {Gr{\"a}fener}, \&
  {Liermann}}]{HGL2006}
{Hamann}, W., {Gr{\"a}fener}, G., \& {Liermann}, A. 2006, \aap, 457, 1015

\bibitem[{{Hamann} \& {Koesterke}(1998)}]{HK1998}
{Hamann}, W. \& {Koesterke}, L. 1998, \aap, 335, 1003

\bibitem[{{Hamann} \& {Gr{\"a}fener}(2003)}]{HG2003}
{Hamann}, W.-R. \& {Gr{\"a}fener}, G. 2003, \aap, 410, 993

\bibitem[{{Hambly} {et~al.}(2008){Hambly}, {Collins}, {Cross}, {Mann}, {Read},
  {Sutorius}, {Bond}, {Bryant}, {Emerson}, {Lawrence}, {Rimoldini}, {Stewart},
  {Williams}, {Adamson}, {Hirst}, {Dye}, \& {Warren}}]{Hambly08}
{Hambly}, N.~C., {Collins}, R.~S., {Cross}, N.~J.~G., {et~al.} 2008, \mnras,
  384, 637

\bibitem[{{Hanson} {et~al.}(1996){Hanson}, {Conti}, \& {Rieke}}]{Hanson96}
{Hanson}, M.~M., {Conti}, P.~S., \& {Rieke}, M.~J. 1996, \apjs, 107, 281

\bibitem[{{Hanson} {et~al.}(2005){Hanson}, {Kudritzki}, {Kenworthy}, {Puls}, \&
  {Tokunaga}}]{Hanson2005}
{Hanson}, M.~M., {Kudritzki}, R.-P., {Kenworthy}, M.~A., {Puls}, J., \&
  {Tokunaga}, A.~T. 2005, \apjs, 161, 154

\bibitem[{{Hanson} {et~al.}(1998){Hanson}, {Rieke}, \& {Luhman}}]{Hanson1998}
{Hanson}, M.~M., {Rieke}, G.~H., \& {Luhman}, K.~L. 1998, \aj, 116, 1915

\bibitem[{{Hewett} {et~al.}(2006){Hewett}, {Warren}, {Leggett}, \&
  {Hodgkin}}]{Hewett06}
{Hewett}, P.~C., {Warren}, S.~J., {Leggett}, S.~K., \& {Hodgkin}, S.~T. 2006,
  \mnras, 367, 454

\bibitem[{{Hodgkin} {et~al.}(2009){Hodgkin}, {Irwin}, {Hewett}, \&
  {Warren}}]{Hodgkin09}
{Hodgkin}, S.~T., {Irwin}, M.~J., {Hewett}, P.~C., \& {Warren}, S.~J. 2009,
  \mnras, 394, 675

\bibitem[{{Langer}(2012)}]{Langer2012}
{Langer}, N. 2012, \araa, 50, 107

\bibitem[{{Lasker} {et~al.}(2008){Lasker}, {Lattanzi}, {McLean}, {Bucciarelli},
  {Drimmel}, {Garcia}, {Greene}, {Guglielmetti}, {Hanley}, {Hawkins},
  {Laidler}, {Loomis}, {Meakes}, {Mignani}, {Morbidelli}, {Morrison},
  {Pannunzio}, {Rosenberg}, {Sarasso}, {Smart}, {Spagna}, {Sturch},
  {Volpicelli}, {White}, {Wolfe}, \& {Zacchei}}]{GSC2008}
{Lasker}, B.~M., {Lattanzi}, M.~G., {McLean}, B.~J., {et~al.} 2008, \aj, 136,
  735

\bibitem[{{Lattimer}(2012)}]{Lattimer2012}
{Lattimer}, J.~M. 2012, in American Institute of Physics Conference Series,
  Vol. 1484, American Institute of Physics Conference Series, ed. S.~{Kubono},
  T.~{Hayakawa}, T.~{Kajino}, H.~{Miyatake}, T.~{Motobayashi}, \& K.~{Nomoto},
  319--326

\bibitem[{{Lawrence} {et~al.}(2007){Lawrence}, {Warren}, {Almaini}, {Edge},
  {Hambly}, {Jameson}, {Lucas}, {Casali}, {Adamson}, {Dye}, {Emerson},
  {Foucaud}, {Hewett}, {Hirst}, {Hodgkin}, {Irwin}, {Lodieu}, {McMahon},
  {Simpson}, {Smail}, {Mortlock}, \& {Folger}}]{Lawrence07}
{Lawrence}, A., {Warren}, S.~J., {Almaini}, O., {et~al.} 2007, \mnras, 379,
  1599

\bibitem[{{Levine} {et~al.}(2004){Levine}, {Rappaport}, {Remillard}, \&
  {Savcheva}}]{Levine04}
{Levine}, A.~M., {Rappaport}, S., {Remillard}, R., \& {Savcheva}, A. 2004,
  \apj, 617, 1284

\bibitem[{{Lewis} {et~al.}(2005){Lewis}, {Irwin}, {Hodgkin}, {Bunclark},
  {Evans}, \& {McMahon}}]{Lewis05}
{Lewis}, J.~R., {Irwin}, M.~J., {Hodgkin}, S.~T., {et~al.} 2005, in
  Astronomical Society of the Pacific Conference Series, Vol. 347, Astronomical
  Data Analysis Software and Systems XIV, ed. P.~{Shopbell}, M.~{Britton}, \&
  R.~{Ebert}, 599

\bibitem[{{Manousakis} {et~al.}(2012){Manousakis}, {Walter}, \&
  {Blondin}}]{Mano2012}
{Manousakis}, A., {Walter}, R., \& {Blondin}, J.~M. 2012, A\&A, 547, A20

\bibitem[{{Morel} \& {Grosdidier}(2005)}]{Morel05}
{Morel}, T. \& {Grosdidier}, Y. 2005, \mnras, 356, 665

\bibitem[{{Oskinova} {et~al.}(2011){Oskinova}, {Todt}, {Ignace}, {Brown},
  {Cassinelli}, \& {Hamann}}]{Oskinova+2011}
{Oskinova}, L.~M., {Todt}, H., {Ignace}, R., {et~al.} 2011, \mnras, 416, 1456

\bibitem[{{Sander} {et~al.}(2012){Sander}, {Hamann}, \& {Todt}}]{SHT2012}
{Sander}, A., {Hamann}, W.-R., \& {Todt}, H. 2012, \aap, 540, A144

\bibitem[{{Sander} {et~al.}(2015){Sander}, {Shenar}, {Hainich},
  {G{\'{\i}}menez-Garc{\'{\i}}a}, {Todt}, \& {Hamann}}]{Sander+2015}
{Sander}, A., {Shenar}, T., {Hainich}, R., {et~al.} 2015, \aap, 577, A13

\bibitem[{{Sim{\'o}n-D{\'{\i}}az} \& {Herrero}(2014)}]{Simon-Diaz2014}
{Sim{\'o}n-D{\'{\i}}az}, S. \& {Herrero}, A. 2014, \aap, 562, A135

\bibitem[{{Skrutskie} {et~al.}(2006){Skrutskie}, {Cutri}, {Stiening},
  {Weinberg}, {Schneider}, {Carpenter}, {Beichman}, {Capps}, {Chester},
  {Elias}, {Huchra}, {Liebert}, {Lonsdale}, {Monet}, {Price}, {Seitzer},
  {Jarrett}, {Kirkpatrick}, {Gizis}, {Howard}, {Evans}, {Fowler}, {Fullmer},
  {Hurt}, {Light}, {Kopan}, {Marsh}, {McCallon}, {Tam}, {Van Dyk}, \&
  {Wheelock}}]{2MASS}
{Skrutskie}, M.~F., {Cutri}, R.~M., {Stiening}, R., {et~al.} 2006, \aj, 131,
  1163

\bibitem[{{Spitzer Science}(2009)}]{IPAC2008}
{Spitzer Science}, C. 2009, VizieR Online Data Catalog, 2293, 0

\bibitem[{{Vink} {et~al.}(2000){Vink}, {de Koter}, \& {Lamers}}]{Vink2000}
{Vink}, J.~S., {de Koter}, A., \& {Lamers}, H.~J.~G.~L.~M. 2000, \aap, 362, 295

\bibitem[{{Walborn}(1976)}]{Walborn1976}
{Walborn}, N.~R. 1976, \apj, 205, 419

\end{thebibliography}

\appendix
\onecolumn

\section{On the clumping factor and mass-loss rate dependency}\label{sec:app}


\begin{figure*}[th]
 \centering
  \includegraphics[width=0.90\textwidth]{normspec}
  \includegraphics[width=0.90\textwidth]{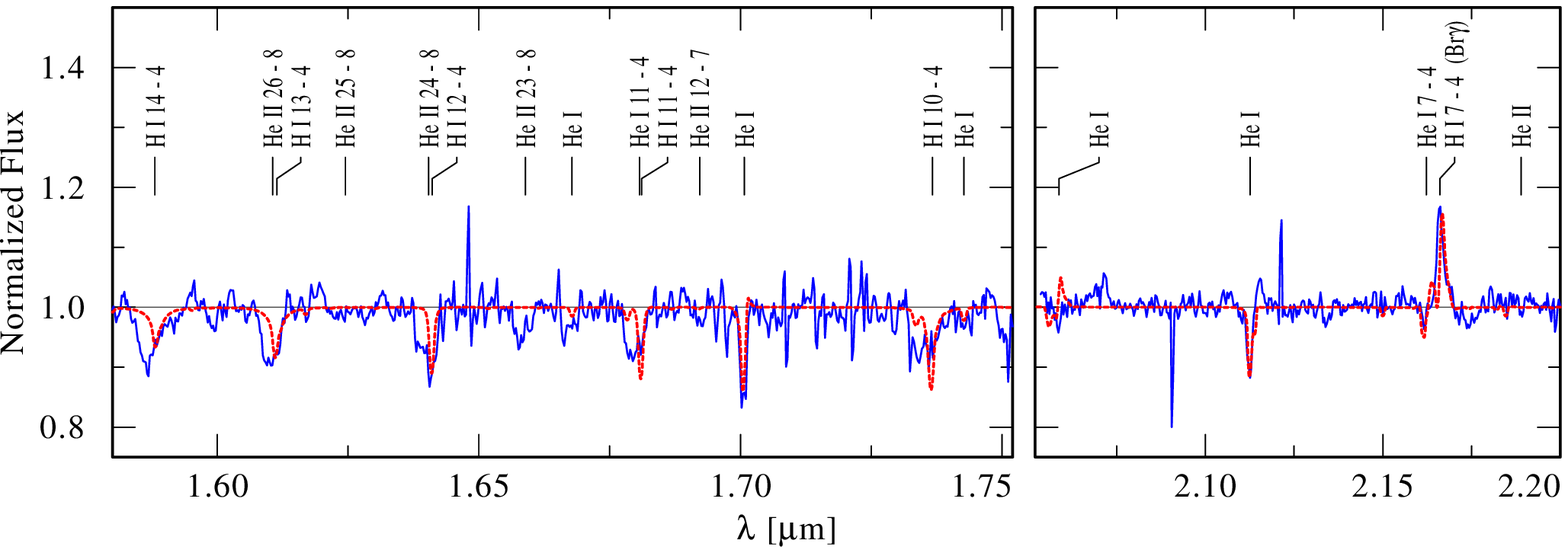}
  \includegraphics[width=0.90\textwidth,angle=360]{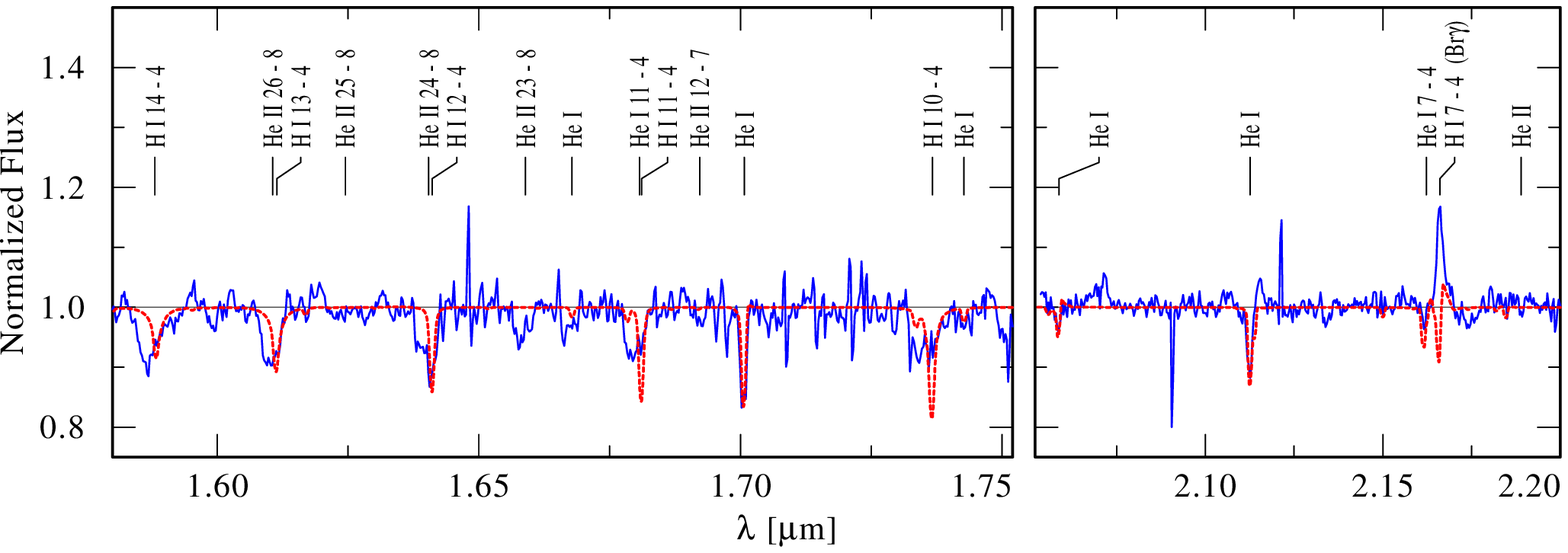}
  \caption{The observed spectrum (thin blue line) -- $H$ band (left panel), $K$ band 
(right panel) -- is compared with the best-fitting PoWR model (dotted line) 
for different D factors: D=10 (top panel); D=4 (middle panel); D=1 (low panel).}
  \label{fig:Dcomp}
\end{figure*}


\end{document}